\documentclass[11pt]{article}
\usepackage[T1]{fontenc}
\usepackage{stix2}
\usepackage{a4wide}
\usepackage{authblk}
\usepackage{amsmath}
\usepackage{graphicx}
\usepackage{float}
\usepackage{overpic}
\usepackage{subcaption}
\usepackage{pgfplots}
\pgfplotsset{compat=1.18}
\usepgfplotslibrary{colormaps,groupplots}
\usepackage{tikz}
\usetikzlibrary{decorations.pathmorphing,calc}
\usepackage{hyperref}
\usepackage{cleveref}

\title{Physics-Informed Neural Operator for Warm-Starting  Background-Decomposed and Preconditioned PSFD: \\ Enabling Scalable 3-D EUV Mask Simulation}
\author{Doyun Kim\thanks{doyun.kim@imec.be}} 
\author{Werner Gillijns\thanks{werner.gillijns@imec.be}}
\affil{IMEC, Kapeldreef 75, 3001 Leuven, Belgium}
\date{}

\begin{document}

\maketitle
\begin{abstract}
    We present a physics-informed neural operator (PINO) trained with pseudo-spectral \\ frequency-domain (PSFD) equations for electromagnetic (EM) scattering problems in EUV lithography. 
    The Fourier neural operator is factorized into a two-dimensional lateral ($xy$) branch and a one-dimensional axial ($z$) branch and is trained self-consistently with background decomposition.
    Thus, the full-vector coupling between the mask and the multilayer response is retained without invoking a finite-order Born approximation.
    In this way, the computational domain size is significantly reduced, thereby lowering the computational cost. 
    The PINO is trained on approximately 16,000 mask designs from the LithoBench library sampled randomly at each training iteration without using precomputed EM field solutions.
    The PINO surrogate model yields predictions with a mean absolute error of about $7 \times 10^{-3}$ for the scattered intensity of held-out mask patterns relative to the reference PSFD solution. 
    Combined with spectral damping, the PINO warm-start initialization accelerates the background-decomposed PSFD solver on finer discretizations.
    
\end{abstract} 

\section{Introduction}
Simulation of electromagnetic fields in three-dimensional structures is a fundamental challenge in computational electromagnetics (EM), with applications spanning photonics, optical metrology, and semiconductor manufacturing\cite{taflove1998, chew1995}. 
In particular, extreme ultraviolet (EUV) lithography has profoundly transformed semiconductor microlithography. 
Achieving sub-nanometer control of printed dimensions over large mask layouts requires electromagnetic simulations that are both highly accurate and computationally efficient, although these objectives are difficult to satisfy simultaneously.
Traditional rigorous electromagnetic solvers, such as finite-difference time-domain (FDTD) and rigorous coupled-wave analysis (RCWA), while accurate, often struggle to meet the computational demands of simulating large-scale EUV mask structures due to their high computational cost and memory requirements \cite{yee1966, moharam1995}. 
For this reason, considerable effort has been devoted to developing
more efficient computational methods that retain the essential physics
while significantly reducing runtime.

A common high-throughput OPC workflow combines a Hopkins-type
partially coherent imaging formulation with a thin-mask complex mask
function, thereby neglecting the volumetric electromagnetic
interactions within the physical mask stack
\cite{hopkins1953,wong2001}.
While this combined approximation is computationally efficient, the
thin-mask representation cannot fully capture the vectorial near-field
interactions and internal coupling that arise in realistic EUV mask
structures, leading to inaccuracies in the predicted aerial images and subsequent lithographic performance\cite{pistor1998, erdmann2006}. 
In EUV lithography, oblique illumination of the reflective mask and
the finite thickness of the absorber stack give rise to pronounced
mask 3D (M3D) effects, including orientation-dependent shadowing and
phase distortions \cite{erdmann2005,pistor2002}.

To overcome these limitations, several computationally efficient EM modeling approaches have been proposed to approximate mask near fields at a lower cost than full rigorous solvers.
Kernel-based compact 3-D mask models approximate rigorous mask near fields by convolving calibrated spatial-domain filters with thin-mask transmission functions, where the filters are calibrated against rigorous electromagnetic simulations of calibration structures
\cite{liu2010mask3d}.
The domain decomposition method (DDM) decomposes complex mask layouts into primitive sub-patterns, computes their EM near-fields rigorously, and assembles the full mask near-field via linear superposition \cite{adam2002}.
Data-driven approaches have also been explored, in which machine learning surrogate models are trained on rigorously computed EM field datasets and applied for fast near-field inference \cite{watanabe2019, jia2014}. 

In particular, physics-informed neural networks and neural operators (PINNs and PINOs) have emerged as powerful tools for solving PDEs and integral equations by embedding the underlying physical laws into the learning process \cite{raissi2019, li2023}. 
Notable advances in PINNs and PINOs have demonstrated their ability to approximate solutions and solution operators constrained by governing equations, including Helmholtz-type formulations relevant to EUV mask diffraction \cite{erdmann2024pinn,medvedev2024,eskin2025}, and to provide efficient surrogates for complex scattering problems \cite{peurifoy2018,khoo2021}.
Nevertheless, scalable three-dimensional EUV mask simulation remains computationally demanding despite advances in both rigorous solvers and approximate mask-modeling approaches \cite{evanschitzky2005euv,zhang2020fast}.

Recently, we presented a pseudo-spectral frequency-domain (PSFD) method featuring a background decomposition approach, where the simulation domain is significantly reduced to only the mask region, and introduced preconditioning techniques that further accelerate convergence \cite{kim2026psfd}. 
In this work, that framework is extended to a PINO based on an
anisotropically factorized Fourier neural operator inspired by F-FNO
\cite{tran2023ffno} and trained with background decomposition.
The resulting surrogate infers 3-D vectorial electric fields for held-out EUV mask geometries without relying on precomputed EM field solutions.
We then examine its transfer to a finer lateral discretization of the
training domain, without retraining, and assess its use as a warm-start
initializer for the background-decomposed PSFD solver with spectral
preconditioning.

\section{Methodology}
\subsection{Factorized Fourier Neural Operator for EM scattering problems}
A neural operator is a deep learning architecture designed to learn mappings between infinite-dimensional function spaces, making it uniquely capable of approximating the solution operators of partial differential equations (PDEs) \cite{li2021,li2023,kovachki2024}. Unlike conventional neural networks, a neural operator processes continuous functions as inputs and outputs. In its foundational formulation, each operator layer typically maps a hidden state via \cite{kovachki2024}
\begin{align}
    \mathcal{N}(u)(x) = \sigma\left( W u(x) + b + \int_{\Omega} K(x,y) u(y) dy \right),
\end{align}
where $u$ is the input function, $\sigma$ is a non-linear activation function, $W$ and $b$ represent a local linear transformation, and $K(x,y)$ denotes a learnable kernel function capturing non-local spatial dependencies. 
To computationally scale this framework to large three-dimensional
configurations, we adopt an anisotropically factorized Fourier neural
operator inspired by the FNO and F-FNO architectures
\cite{li2021,tran2023ffno}.
Instead of computing a single fully coupled 3-D spectral kernel, each layer is decomposed into a sequential composition of a 2-D Fourier layer acting on independent $xy$-slices, a 1-D Fourier layer along the axial ($z$) direction, and a pointwise channel-mixing branch.

More explicitly, let $v^{(\ell)}$ denote the hidden feature at layer $\ell$. The lateral ($xy$) spectral branch applies a two-dimensional Fourier operator independently on each $z$-slice:
\begin{align}
    u^{(\mathrm{xy},\ell)}_{o}(x,y,z)
    =
    \mathcal{F}_{2}^{-1}\!\left[
        \sum_{i=1}^{d_v}
        \mathcal{K}^{\mathrm{xy},(\ell)}_{oi}(k_x,k_y)\,
        \mathcal{F}_{2}
        \big(v^{(\ell)}_{i}(\cdot,\cdot,z)\big)
    \right].
\end{align}
The lateral kernel $\mathcal{K}^{\mathrm{xy},(\ell)}_{oi}(k_x,k_y)$ is nonzero only for $|k_x|\le k_x^{\max}$ and $|k_y|\le k_y^{\max}$, corresponding to a hard spectral truncation at the retained Fourier modes.

The axial ($z$) branch is then applied to the laterally filtered feature:
\begin{align}
    u^{(\mathrm{z},\ell)}_{o}(x,y,z)
    =
    \mathcal{F}_{1}^{-1}\!\left[
        \sum_{i=1}^{d_v}
        \mathcal{K}^{\mathrm{z},(\ell)}_{oi}(k_z)\,
        \mathcal{F}_{1}
        \big(u^{(\mathrm{xy},\ell)}_{i}(x,y,\cdot)\big)
    \right],
\end{align}
where $\mathcal{K}^{\mathrm{z},(\ell)}_{oi}(k_z)$ is nonzero only for $|k_z|\le k_z^{\max}$.

Finally, these filtered features are combined and updated via a local pointwise multilayer perceptron (MLP) with a residual connection:
\begin{align}
    v^{(\ell+1)}_{o}(\mathbf{x})
    =
    v^{(\ell)}_{o}(\mathbf{x})
    +
    \sigma_{\ell}\!\left(
        \sum_{j=1}^{d_h}
        W^{(\ell)}_{2,oj}
        \sigma\!\left(
            \sum_{i=1}^{d_v}
            W^{(\ell)}_{1,ji}
            \left(
                u^{(\mathrm{xy},\ell)}_{i}(\mathbf{x})
                +
                u^{(\mathrm{z},\ell)}_{i}(\mathbf{x})
            \right)
            +
            b^{(\ell)}_{1,j}
        \right)
        +
        b^{(\ell)}_{2,o}
    \right).
\end{align}
Here, $\sigma$ denotes the GeLU activation. GeLU is also applied after the pointwise MLP in all intermediate layers, whereas this outer activation is omitted in the final layer.
The matrices $W_{1}^{(\ell)}$ and $W_{2}^{(\ell)}$, together with the bias vectors $b_{1}^{(\ell)}$ and $b_{2}^{(\ell)}$, are the learnable parameters of the pointwise MLP at layer $\ell$.
The indices $i$, $j$, and $o$ denote the input, hidden, and output channels, respectively. 
Here, $d_v$ is the number of feature channels in the Fourier layer and $d_h$ is the hidden-channel dimension of the MLP.

By composing $L$ such operator layers, the network iteratively extracts and refines the non-local spatial interactions inherent to the electromagnetic scattering process. 
To initialize the forward evaluation, the mask geometry is encoded by a binary material-indicator field $a_{\mathrm{bin}}(\mathbf{x})\in\{0,1\}$. 
Before network evaluation, this indicator is Gaussian-smoothed on the computational grid to produce a regularized input field $a(\mathbf{x})\in[0,1]$. 
The regularized input field is then fed into the PINO as its structural input.
This input field, concatenated with the grid coordinates $(x,y,z)$ as additional channels, is first lifted to a higher-dimensional channel space via a shallow pointwise linear transformation, commonly referred to as a lifting layer, to formulate the initial hidden state $v^{(0)}(\mathbf{x})$. Following the sequential application of the $L$ factorized Fourier layers, the final hidden representation $v^{(L)}(\mathbf{x})$ is projected back to the physical target space through a local decoding layer. 

Consequently, the complete end-to-end architecture acts as a parameterized surrogate operator $\mathcal{G}_{\theta}$, where $\theta$ denotes the set of all learnable weights and biases. 
Ultimately, the network maps the regularized input field to the
predicted 3-D vectorial scattered electric field, expressed as
\begin{align}
    \mathcal{G}_{\theta} : a(\mathbf{x}) \mapsto \widehat{\mathbf{E}}_{\mathrm{scat}}(\mathbf{x}) \in \mathbb{C}^3.
\end{align}
The optimal parameters $\theta$ are subsequently determined by minimizing a designated loss function, thereby training the network to accurately approximate the true solution operator.

\subsection{Loss function for physics-informed neural operator}
The loss function of the PINO surrogate model is constructed using the PSFD equation with the background decomposition method~\cite{kim2026psfd}.
The detailed formulation can be found in our previous paper; here, we briefly summarize the key points. 

At every training iteration, the model predicts the scattered field $\widehat{\mathbf{E}}_{\theta, \mathrm{scat}}$ strictly within the mask scatterer domain. Given this prediction, the reflected background field $\mathbf{E}_{\mathrm{ML}}(\mathbf{r})$ is reconstructed from the total field at the computationally defined multilayered interface $z=z_{\mathrm{ml}}$. 
Specifically, we define a reflection operator $\mathcal{R}_{\mathrm{ml}}$ based on the transfer-matrix method (TMM) that maps the transverse field components at the interface $z=z_{\mathrm{ml}}$ to the reflected background field, expressed as the functional mapping:
\begin{align}
    \mathcal{R}_{\mathrm{ml}} :
    \left[
    \widehat{\mathbf{E}}_{\theta,\mathrm{scat}}(\mathbf{r})
    +
    \mathbf{E}_{\mathrm{inc}}(\mathbf{r})
    \right]_{\parallel,\,z=z_{\mathrm{ml}}}
    \mapsto
    \mathbf{E}_{\mathrm{ML}}(\mathbf{r}).
\end{align}
This operator inherently incorporates (i) upgoing/downgoing field decomposition, (ii) TE/TM mode decomposition, (iii) the TMM-based reflection response at the fictitious interface, and (iv) an analytical phase-matching propagation to restore the full 3D spatial field distribution $\mathbf{E}_{\mathrm{ML}}(\mathbf{r})$. This reconstructed field is added to the incident field $\mathbf{E}_{\mathrm{inc}}$ to update the total background field. 
By self-consistently coupling the fields in the mask-only domain to the analytically evaluated multilayer response at every iteration, the formulation retains the full-vector Maxwell coupling between the mask and the planar multilayer without invoking a finite-order Born
approximation. 
Remaining errors arise from finite spatial
discretization, PML truncation, interface regularization, iterative solver tolerances, and neural-operator approximation.

The physics-informed loss function $\mathcal{J}(\theta)$ is formulated by defining an effective source term, representing the total background fields including the incident field, as:
\begin{align}
    \mathbf{b}_{\mathrm{bg}}(\mathbf{r})
    =
    k_0^2\,\Delta\varepsilon(\mathbf{r})
    \Big(
        \mathbf{E}_{\mathrm{inc}}(\mathbf{r})
        +
        \mathbf{E}_{\mathrm{ML}}(\mathbf{r})
    \Big).
    \label{eq:bg_source}
\end{align}
Here,
$\Delta\varepsilon(\mathbf{r})
=
\varepsilon(\mathbf{r})
-
\varepsilon_{\mathrm{bg}}(\mathbf{r})$.

Consequently, the mean-squared residual of the governing vectorial
PSFD equation is minimized together with a frequency-weighted penalty
that emphasizes higher spatial-frequency components:
\begin{align}
    \mathcal{J}(\theta)
    &=
    \frac{1}{N}
    \sum_{\alpha \in \{x,y,z\}}
    \left\|
        \left[
        \mathcal{L}\,
        \widehat{\mathbf{E}}_{\theta,\mathrm{scat}}
        -
        \mathbf{b}_{\mathrm{bg}}
        \right]_{\alpha}
    \right\|_2^2
    \nonumber\\
    &\quad+
    \frac{\lambda_{\mathrm{hf}}}{N}
    \sum_{\alpha \in \{x,y,z\}}
    \left\|
        w(\boldsymbol{\nu})\,
        \mathcal{F}_{3}
        \left[
            \left(
            \mathcal{L}\,
            \widehat{\mathbf{E}}_{\theta,\mathrm{scat}}
            -
            \mathbf{b}_{\mathrm{bg}}
            \right)_{\alpha}
        \right]
    \right\|_2^2,
    \label{eq:loss}
\end{align}
where $N=N_xN_yN_z$ is the number of spatial grid points,
$\alpha$ denotes the Cartesian field component,
$\mathcal{F}_{3}$ is the three-dimensional discrete Fourier transform, and
\begin{align}
    w(\boldsymbol{\nu})
    =
    w_0+\eta_{\mathrm{hf}}|\boldsymbol{\nu}|,
    \qquad
    |\boldsymbol{\nu}|
    =
    \sqrt{\nu_x^2+\nu_y^2+\nu_z^2}.
\end{align}
is a frequency-dependent weight that increasingly emphasizes residual
components at higher spatial frequencies.
Here, $\lambda_{\mathrm{hf}}$ controls the contribution of the
frequency-weighted penalty, $w_0$ is the baseline weight, and
$\eta_{\mathrm{hf}}$ determines the strength of the high-frequency
emphasis.
The operator $\mathcal{L}$ is the vectorial Helmholtz operator given by
\begin{align}
    \mathcal{L}
    &=
    \nabla \times \nabla \times
    -
    k_{0}^{2}\varepsilon(\mathbf{r})\mathbf{I}
    \nonumber\\
    &=
    \begin{bmatrix}
        -\partial_y\partial_y-\partial_z\partial_z-k_0^2\varepsilon
        &
        \partial_x\partial_y
        &
        \partial_x\partial_z
        \\
        \partial_y\partial_x
        &
        -\partial_x\partial_x-\partial_z\partial_z-k_0^2\varepsilon
        &
        \partial_y\partial_z
        \\
        \partial_z\partial_x
        &
        \partial_z\partial_y
        &
        -\partial_x\partial_x-\partial_y\partial_y-k_0^2\varepsilon
    \end{bmatrix}.
    \label{eq:vector_helmholtz_operator}
\end{align}

Here, $\partial_\alpha$ denotes differentiation with respect to the
coordinate $\alpha$. In the numerical implementation, all derivatives,
including the mixed terms, are evaluated pseudo-spectrally with PML
coordinate stretching as detailed in Ref.~\cite{kim2026psfd}.
Precomputed simulation results are not used for training the PINO
surrogate model.

\section{Results}
In this section, we present numerical results demonstrating the accuracy of the background-decomposed PSFD formulation, the inference accuracy of the trained PINO surrogate, its scalability to simulation domains at resolutions finer than those used during training, and the convergence acceleration achieved through PINO warm-start initialization. 
Throughout all simulations, the mask structure is illuminated by a TE-polarized ($E_y$) plane wave at the EUV wavelength $\lambda = 13.5\,\mathrm{nm}$ under normal incidence. Following Ref.~\cite{makhotkin2021}, the multilayer substrate is modeled as a Mo/Si distributed Bragg reflector with 40 bilayer pairs, each with a nominal period of $7\,\mathrm{nm}$, for a total stack thickness of approximately $280\,\mathrm{nm}$.

\paragraph*{Background-decomposition.}
We first show the accuracy advantage of the background-\\decomposed PSFD formulation~\cite{kim2026psfd} over a standard full-domain PSFD solver that numerically discretizes the entire multilayer stack. 
This accuracy advantage carries over naturally to the PINO surrogate, since its physics-informed objective is constructed from the background-decomposed PSFD residual. 
If the PINO were instead trained using the residual of a full-domain PSFD solver, systematic multilayer discretization errors would become entangled with neural-operator approximation errors.


\begin{figure}[htbp]
    \centering
    \begin{tikzpicture}
        \node[anchor=south west, inner sep=0] (image) at (0,0)
            {\includegraphics[width=1.0\textwidth]{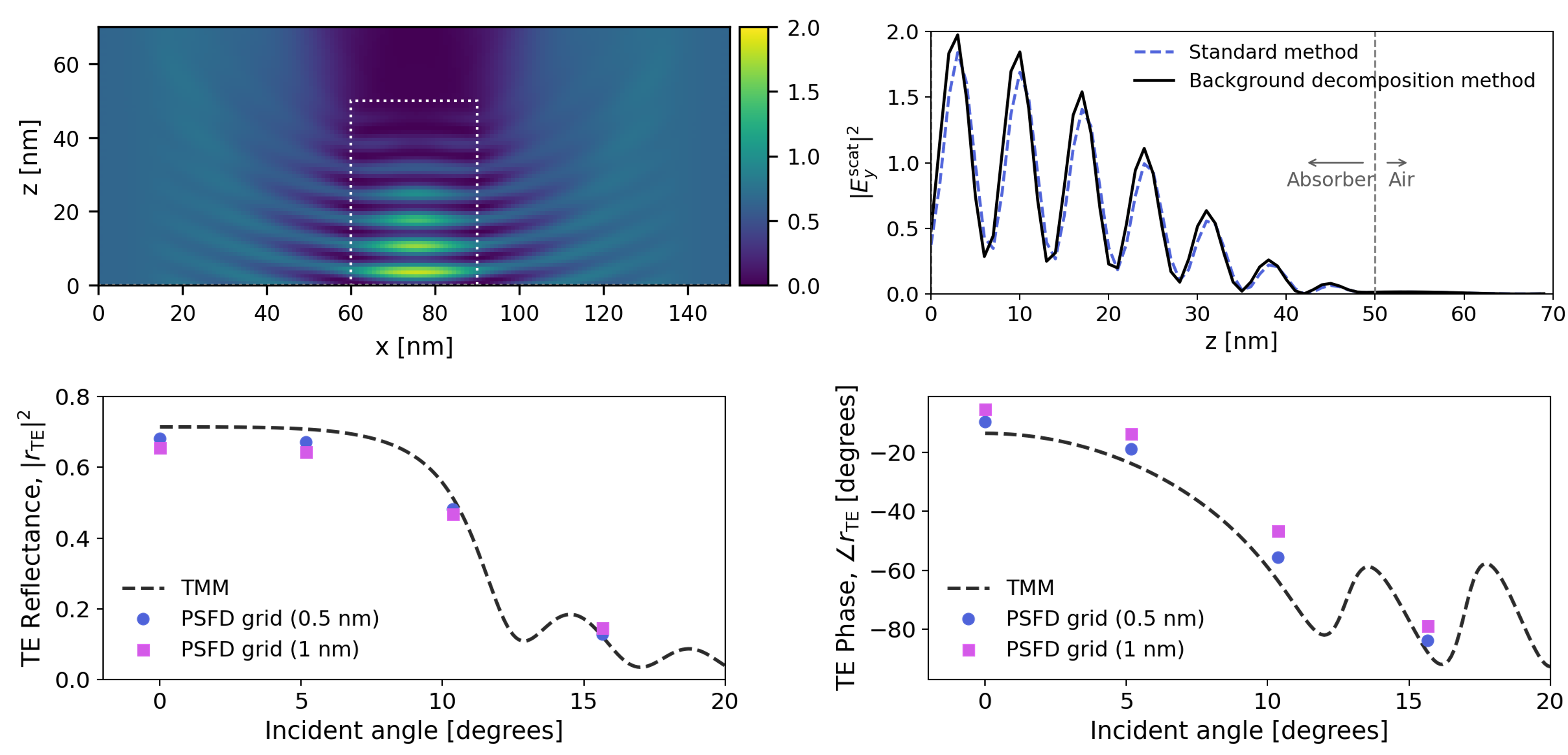}};
        \begin{scope}[shift={(image.south west)},
                      x={(image.south east)}, y={(image.north west)}]
            \node[anchor=north west, font=\bfseries, inner sep=1.5pt]
                at (0.01, 1.05) {(a)};
            \node[anchor=north west, font=\bfseries, inner sep=1.5pt]
                at (0.55, 1.05) {(b)};
            \node[anchor=north west, font=\bfseries, inner sep=1.5pt]
                at (0.01, 0.55) {(c)};
            \node[anchor=north west, font=\bfseries, inner sep=1.5pt]
                at (0.55, 0.55) {(d)};
        \end{scope}
    \end{tikzpicture}
    \caption{%
        (a) Scattered intensity $|E_y^{\mathrm{scat}}|^2$ computed by the background-decomposed PSFD solver with the computational domain covering only the 50\,nm Ta absorber region (dashed outline),
        (b) Vertical profile of $|E_y^{\mathrm{scat}}|^2$ along the $z$-axis at the center of the absorber region for both solvers at $\Delta x = 1\,$nm,
        (c, d) TE reflectance $|r_{\mathrm{TE}}|^2$ and reflection phase $\arg(r_{\mathrm{TE}})$ of the bare multilayer computed by the standard solver at $\Delta x = 1\,$nm and $0.5\,$nm, compared with the exact TMM result. 
    }
    \label{fig:figure1}
\end{figure}
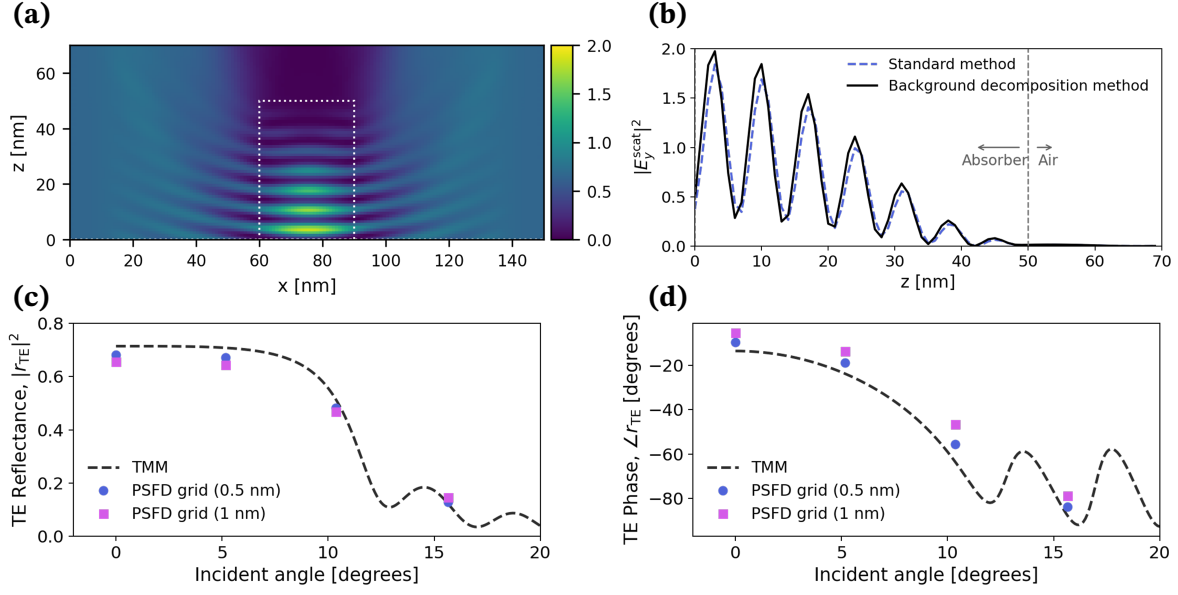

\Cref{fig:figure1}(a) shows the scattered intensity $|E_y^{\mathrm{scat}}|^2$ obtained by the background-decomposed PSFD solver within the computational domain spanning only the $50\,\mathrm{nm}$-thick Ta absorber region (dashed outline), with the $280\,\mathrm{nm}$-thick underlying multilayer stack excluded from explicit discretization. The multilayer-reflected background field, $\mathbf{E}_{\mathrm{ML}}(\mathbf{r})$, is updated at every iteration via TMM and imposed as a self-consistent reflected-background contribution, as defined in Eq.~\eqref{eq:bg_source}, while the iterative PSFD solver resolves only the absorber-induced scattered field $\mathbf{E}^{\mathrm{scat}}$. Removing the multilayer from the computational domain reduces the domain height substantially, leading to a corresponding decrease in both memory overhead and the number of solver iterations. The reduction in vertical ($z$-axis) grid points can be redistributed to enlarge the in-plane ($xy$-plane) simulation area, directly enabling large-area lateral scaling at a fixed memory budget.

\Cref{fig:figure1}(b) compares the line profile of the scattered intensity $|E_y^{\mathrm{scat}}|^2$ extracted at $x = 75\,\mathrm{nm}$ from the 2D field shown in \Cref{fig:figure1}(a), as obtained by the background-decomposed PSFD solver (black line) and a standard PSFD solver that explicitly discretizes the entire multilayer stack (dotted blue line), both computed at a uniform grid spacing of $\Delta x = \Delta y = \Delta z = 1\,\mathrm{nm}$. A measurable discrepancy in $|E_y^{\mathrm{scat}}|^2$ is observed between the two methods along the entire profile. This discrepancy is attributed to a discretization error that arises when numerically resolving the optically thick multilayer. The complex interference between upward- and downward-propagating modes within the stack is sensitive to the grid spacing. A finite grid spacing introduces a grid-size-dependent phase error that causes a fraction of the reflected scattered energy to leak numerically rather than being faithfully returned to the absorber region.

\Cref{fig:figure1}(c) and 1(d) show the TE-polarized reflectance $|r_{\mathrm{TE}}|^2$ and the reflection phase $\arg(r_{\mathrm{TE}})$ of the bare multilayer stack without the absorber, as computed by the standard solver at grid spacings $\Delta x = 1\,\mathrm{nm}$ and $\Delta x = 0.5\,\mathrm{nm}$. Both quantities are compared with the analytically exact TMM results to quantify the numerical leakage introduced by discretizing the multilayer stack. With discrete grid spacing $\Delta x$, the transverse wavenumbers supported by the Fourier grid are restricted to $k_{x,m} = 2\pi m/(N_x\Delta x)$, so that only a discrete set of incidence angles $\theta_m = \arcsin(k_{x,m}/k_0)$ can be faithfully represented without inducing angular dispersion artifacts. For the two grid spacings considered here, the representable angles within the range $0^{\circ}$--$20^{\circ}$ are $\theta = 0^{\circ}$, $5.16^{\circ}$, $10.37^{\circ}$, and $15.66^{\circ}$, so the comparison is accordingly restricted to these angles. Both grid spacings yield $|r_{\mathrm{TE}}|^2$ and $\arg(r_{\mathrm{TE}})$ values that systematically deviate from the TMM references, with the deviation decreasing as $\Delta x$ is reduced. 
The systematic decrease in the deviation as $\Delta x$ is reduced is consistent with a multilayer-discretization error that converges under grid refinement.
Achieving this limit requires increasing the number of grid points $N_x$ at a fixed physical domain size, which scales the memory requirement and imposes a prohibitive memory overhead.
Thus, the background-decomposed PSFD solver avoids the multilayer-discretization error and memory overhead associated with explicitly meshing the stack. 

\paragraph*{PINO surrogate.}
The PINO is trained on a fixed dataset of approximately 16,000 mask designs drawn from the LithoBench benchmark library, with mini-batches sampled randomly at each training iteration.
For inference, the trained PINO is evaluated on a separate set of 100 held-out mask designs that were not included in the training dataset.
Each sample is defined on a computational domain of
$600\times600\times200\,\mathrm{nm}^3$ at grid spacings of
$\Delta x=\Delta y=4\,\mathrm{nm}$ and
$\Delta z=2\,\mathrm{nm}$.


\begin{figure}[htbp]
    \centering
    \begin{tikzpicture}
        \node[anchor=south west, inner sep=0] (image) at (0,0)
            {\includegraphics[width=1.0\textwidth]{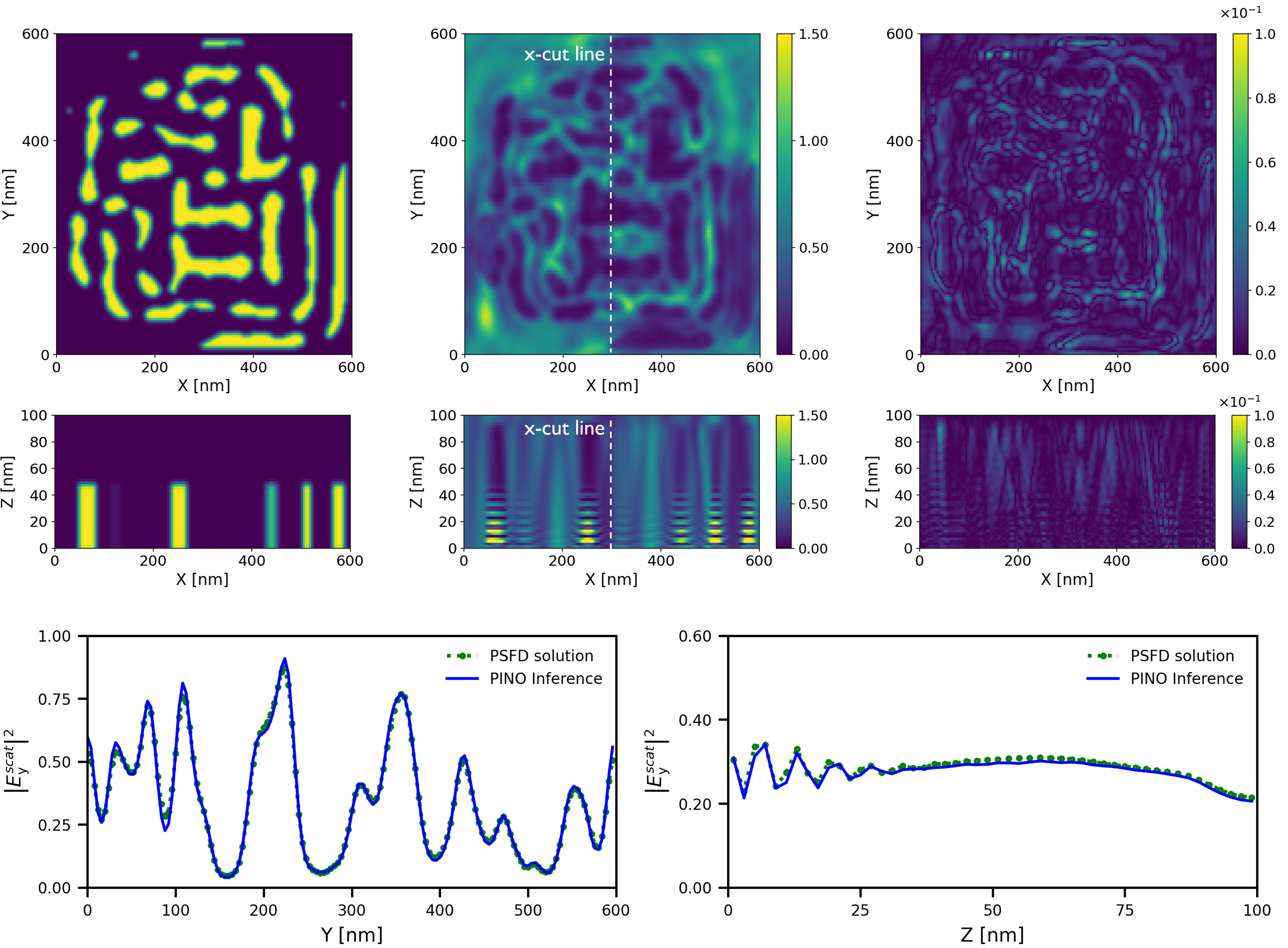}};
        \begin{scope}[shift={(image.south west)},
                      x={(image.south east)}, y={(image.north west)}]
            \node[anchor=north west, font=\bfseries, inner sep=1.5pt]
                at (0.01, 1.02) {(a)};
            \node[anchor=north west, font=\bfseries, inner sep=1.5pt]
                at (0.34, 1.02) {(b)};
            \node[anchor=north west, font=\bfseries, inner sep=1.5pt]
                at (0.67, 1.02) {(c)};
            \node[anchor=north west, font=\bfseries, inner sep=1.5pt]
                at (0.01, 0.38) {(d)};
            \node[anchor=north west, font=\bfseries, inner sep=1.5pt]
                at (0.51, 0.38) {(e)};
        \end{scope}
    \end{tikzpicture}
    \caption{%
        (a) Regularized input field encoding the material distribution of a representative test mask, shown as horizontal ($xy$-plane) and vertical ($xz$-plane) cross-sections.
        (b) Scattered intensity $|E_y^{\mathrm{scat}}|^2$ just above the
        absorber as inferred by the trained PINO surrogate, shown in the
        $xy$- and $xz$-planes.
        (c) Mean absolute error of the PINO inference with respect to the background-decomposed PSFD reference, shown in the $xy$- and $xz$-planes. 
        (d, e) One-dimensional profiles of $|E_y^{\mathrm{scat}}|^2$ along
        the $y$-axis and $z$-axis at the $x$-cut indicated in (b), comparing
        the PINO inference with the background-decomposed PSFD reference.
    }
    \label{fig:figure2}
\end{figure}
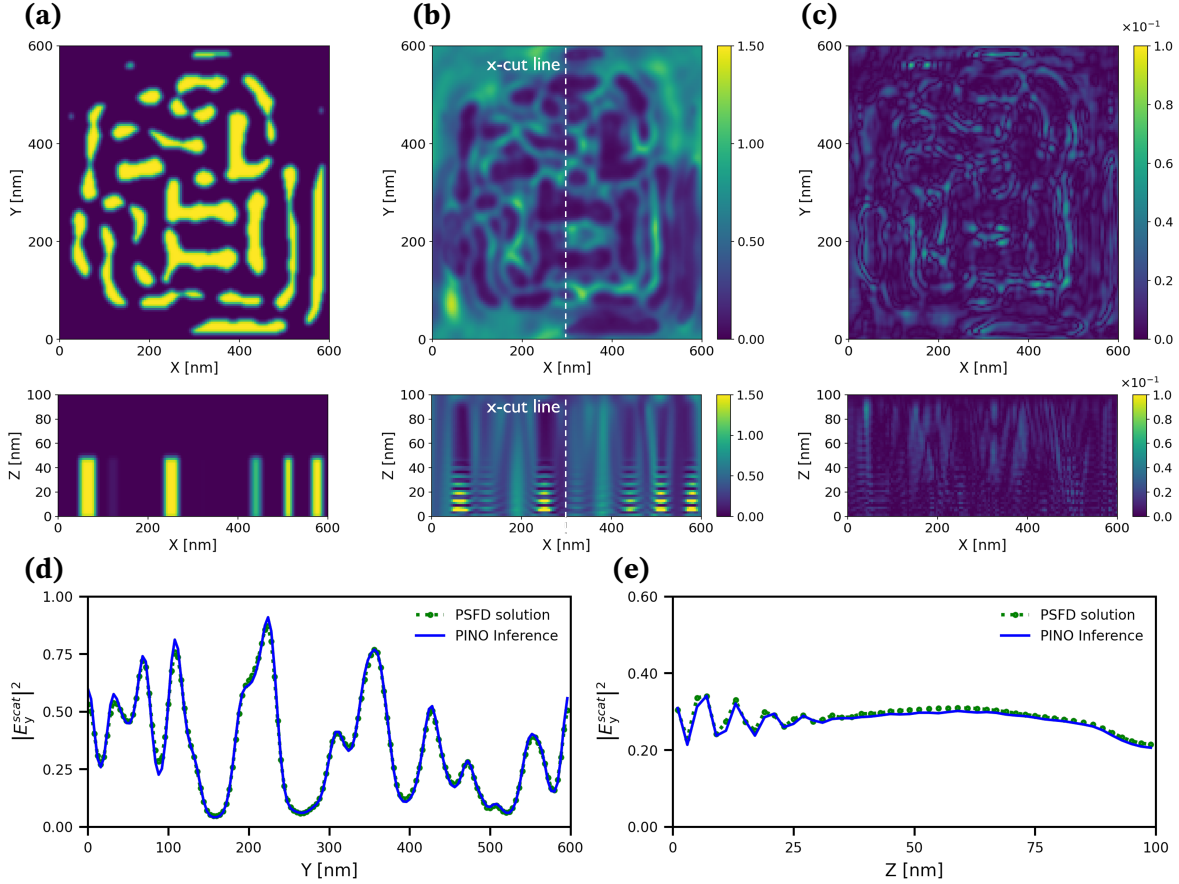

\Cref{fig:figure2}(a) shows horizontal ($xy$-plane) and vertical ($xz$-plane) cross-sections of the regularized input field for a representative test mask.
Since a discontinuous binary interface contains spatial-frequency components beyond those representable on a finite Fourier grid, its direct truncation produces Gibbs ringing and can degrade convergence during training with the spectral loss.
A Gaussian smoothing is therefore applied to the binary indicator \cite{fornberg1998,liu1997,farjadpour2006} to obtain the regularized input field.
The same Gaussian smoothing is applied to the structure used by the rigorous PSFD solver before constructing the material distribution.
\Cref{fig:figure2}(b) shows the scattered intensity $|E_y^{\mathrm{scat}}|^2$ evaluated just above the absorber, as predicted by the trained PINO surrogate. \Cref{fig:figure2}(c) shows the corresponding mean absolute error (MAE).
All pointwise errors and MAEs reported in this work are computed from absolute differences between the PINO-predicted and PSFD-reference scattered intensities $|E_y^{\mathrm{scat}}|^2$.
The PINO is trained on a single NVIDIA A100 80 GB GPU until convergence, at which point the physics residual reaches approximately $10^{-2}$. The correspondence with the background-decomposed PSFD solver is more clearly quantified in \Cref{fig:figure2}(d) and 2(e), which compare one-dimensional profiles of $|E_y^{\mathrm{scat}}|^2$ extracted along the $y$-axis in the $xy$-plane and along the $z$-axis in the $xz$-plane, respectively. 
Both profiles show excellent agreement between the PINO prediction and the PSFD reference, with an MAE of approximately $7 \times 10^{-3}$.

\begin{figure}[H]
    \centering
    \includegraphics[width=0.7\textwidth]{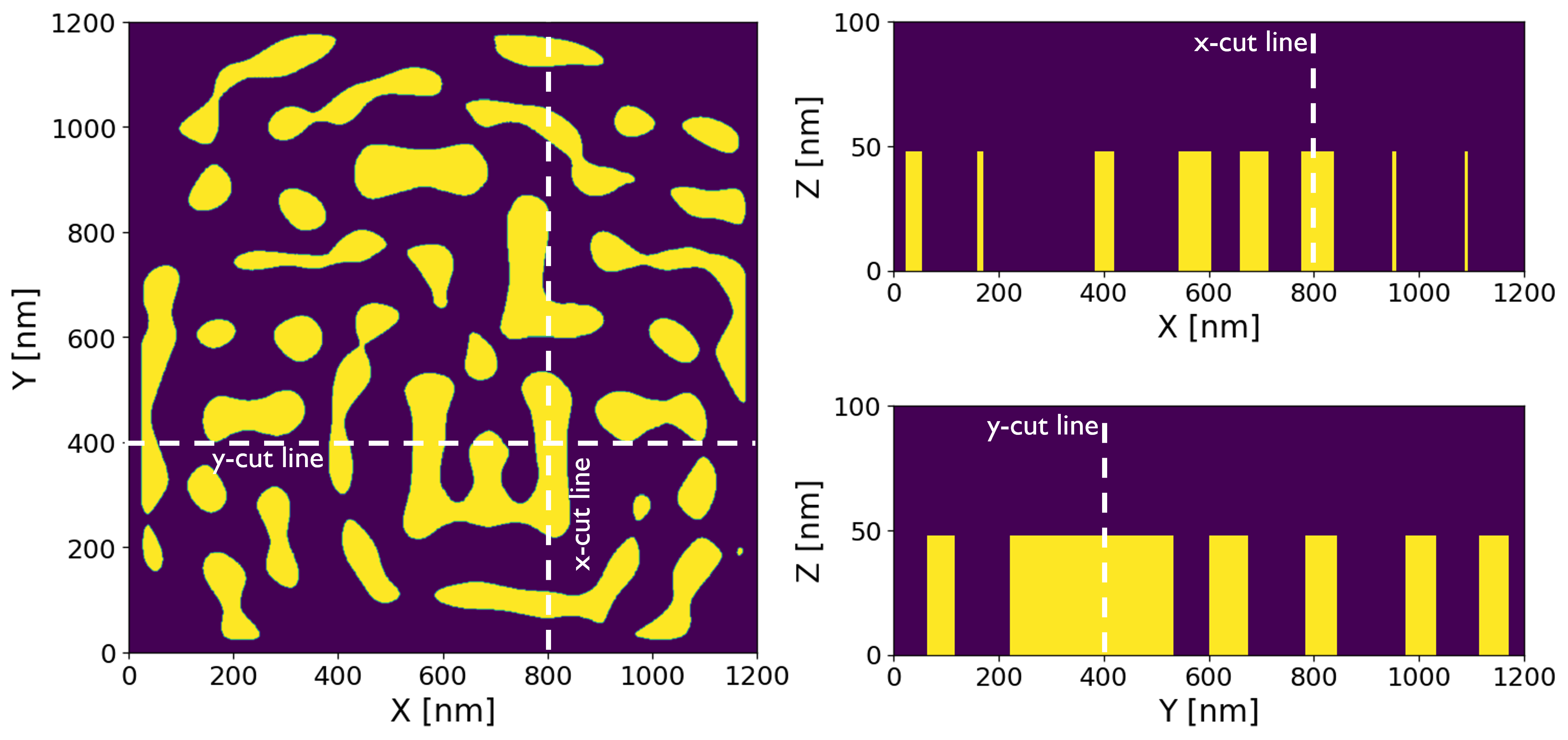}
    \caption{%
        Regularized input field of a representative held-out mask gemetry, displayed as two-dimensional cross-sections in the $xy$-, $xz$-, and $yz$-planes.
    }
    \label{fig:figure3}
\end{figure}

\paragraph*{Application to finer inference grids.}
We next examine the transfer of the trained PINO surrogate to an inference grid finer than that used during training. In this study, the target inference grid is $N_x \times N_y \times N_z = 1200 \times 1200 \times 100$ at $\Delta x = \Delta y = 1\,\mathrm{nm}$ and $\Delta z = 2\,\mathrm{nm}$, corresponding to the same physical domain of $L_x \times L_y \times L_z = 1200 \times 1200 \times 200\,\mathrm{nm}^3$. 
As a test case for this evaluation, \Cref{fig:figure3} shows a representative regularized mask design from the test set, displayed as two-dimensional cross-sections in the $xy$-, $xz$-, and $yz$-planes. 


The dashed lines indicate the cut positions used to extract the $xy$- and $yz$-plane cross-sections, and the $z$-cut is taken at the center of the absorber layer. 
Direct training of the FNO on the target inference grid of $N_x \times N_y \times N_z = 1200 \times 1200 \times 100$ is computationally challenging, since training requires simultaneously holding in device memory the intermediate layer activations for backpropagation, the gradient tensors, and the Adam optimizer state buffers, all of which scale as $\mathcal{O}(N_xN_yN_z)$. None of these are required during inference~\cite{li2021,li2023,li2023gino}. The training bottleneck therefore imposes a practical upper limit on the domain size under given GPU memory constraints. 
To remain within the GPU memory budget, the PINO is trained on a coarser grid of $N_x \times N_y \times N_z = 300 \times 300 \times 100$ at grid spacings of $\Delta x = \Delta y = 4\,\mathrm{nm}$ and $\Delta z = 2\,\mathrm{nm}$, corresponding to a physical domain of $L_x \times L_y \times L_z = 1200 \times 1200 \times 200\,\mathrm{nm}^3$. Inference is then performed at finer grid spacings of $\Delta x = \Delta y = 1\,\mathrm{nm}$ and $\Delta z = 2\,\mathrm{nm}$, increasing the grid resolution to $N_x \times N_y \times N_z = 1200 \times 1200 \times 100$.

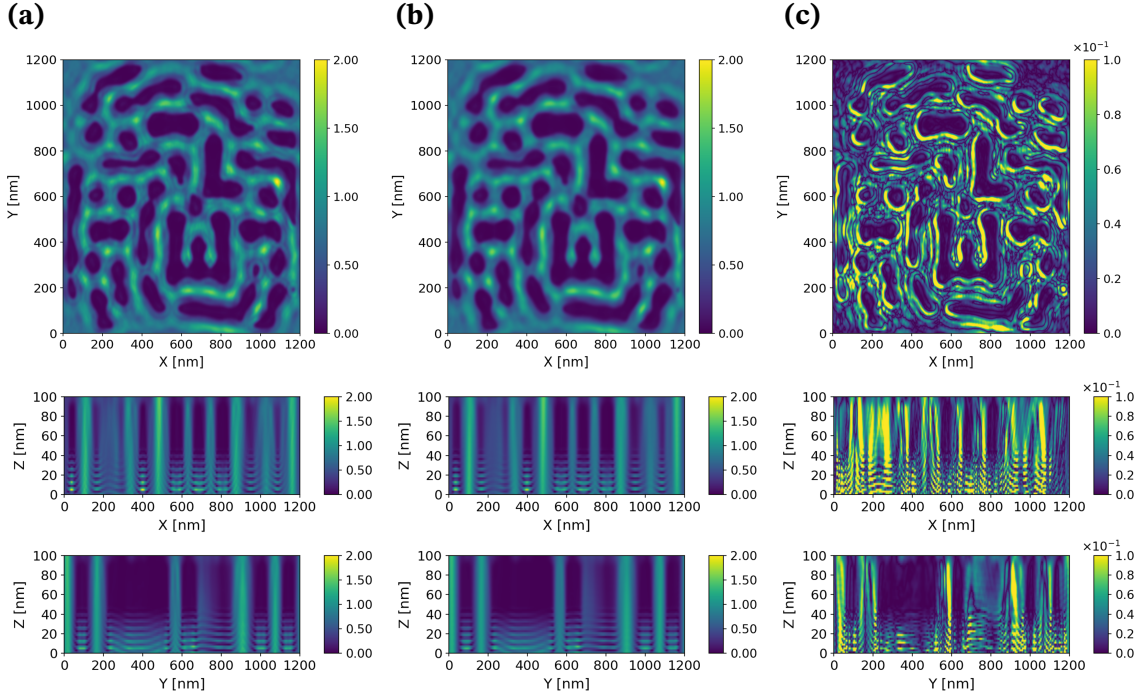
\begin{figure}[H]
    \centering
    \begin{tikzpicture}
        \node[anchor=south west, inner sep=0] (image) at (0,0)
            {\includegraphics[width=1.0\textwidth]{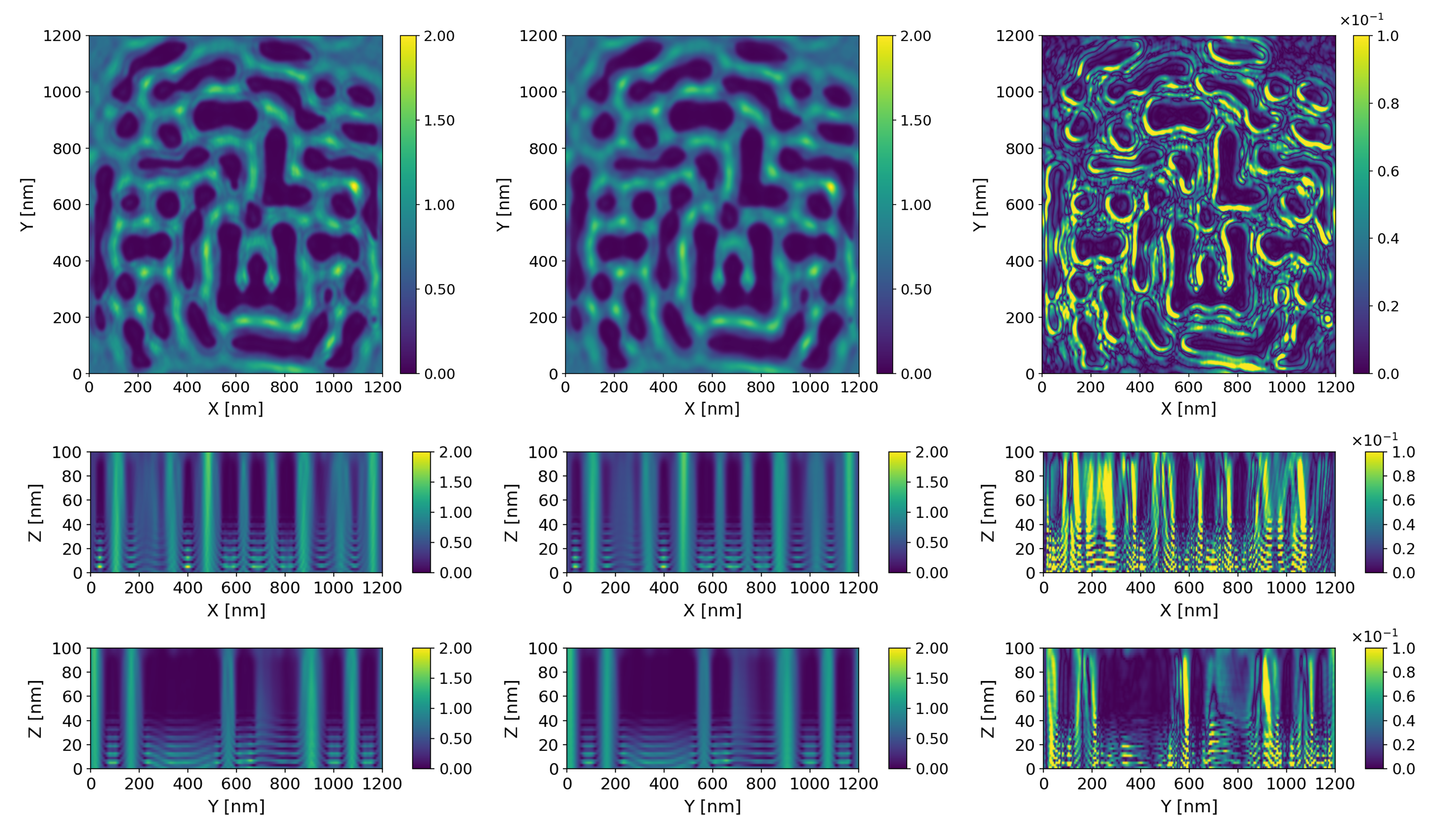}};
        \begin{scope}[shift={(image.south west)},
                      x={(image.south east)}, y={(image.north west)}]
            \node[anchor=north west, font=\bfseries, inner sep=1.5pt]
                at (0.01, 1.05) {(a)};
            \node[anchor=north west, font=\bfseries, inner sep=1.5pt]
                at (0.34, 1.05) {(b)};
            \node[anchor=north west, font=\bfseries, inner sep=1.5pt]
                at (0.67, 1.05) {(c)};
        \end{scope}
    \end{tikzpicture}
    \caption{%
        Comparison of the scattered intensity $|E_y^{\mathrm{scat}}|^2$ for the test mask of Fig.~\ref{fig:figure3} at $\Delta x = \Delta y = 1\,$nm, four times finer than the training grid spacing.
        (a) background-decomposed PSFD reference, (b) PINO inference, and (c) pointwise absolute difference between (a) and (b).
    }
    \label{fig:figure4}
\end{figure}

The scattered intensity $|E_y^{\mathrm{scat}}|^2$ computed for this mask geometry is shown in \Cref{fig:figure4}, where \Cref{fig:figure4}(a) presents the PINO inference and \Cref{fig:figure4}(b) the corresponding background-decomposed PSFD reference at $\Delta x = \Delta y = 1\,\mathrm{nm}$, with their pointwise absolute difference given in \Cref{fig:figure4}(c).
The quantitative comparison yields an MAE of approximately $3 \times 10^{-2}$.
These values are roughly half an order of magnitude larger than the corresponding errors observed in \Cref{fig:figure2}, where the grid spacings are matched.
The errors at the edges of the geometries are more pronounced as seen in \Cref{fig:figure2}(c) and \Cref{fig:figure4}(c).
The difference can be attributed to higher spatial frequencies supported by the finer inference grid but not represented on the coarser training grid.


\paragraph*{PINO warm-start for preconditioned PSFD with Krylov iteration.}
We next investigate the trained PINO model as a warm-start initializer for the preconditioned Krylov solver underlying the iterative background-decomposed PSFD scheme.
This approach is designed to improve the early-stage convergence behavior and reduce the number of iterations to meet practical convergence criteria. 
The linear system arising from the background-decomposed PSFD discretization is solved using the BiCGSTAB algorithm~\cite{saad2003,vandervorst1992}, which is a Krylov subspace method well-suited to complex, non-Hermitian systems of the form encountered in frequency-domain electromagnetic scattering.

Right preconditioning~\cite{benzi2002,saad2003} introduces an auxiliary variable for the Krylov iteration. 
With this approach, convergence can be improved by transforming the original system into a more favorable spectral form through a preconditioner that approximates selected features of the inverse system matrix $\mathbf{A}^{-1}$.
In addition, initialization~\cite{saad2003} supplies a starting iterate $\mathbf{E}^{(0)}$ that reduces the initial residual and shortens the early transient correction phase before the asymptotic regime is reached. 
In this work, three configurations are compared: (i) null preconditioner and null initialization, (ii) the spectral damping preconditioner~\cite{kim2026psfd} and null initialization, and (iii) the spectral damping preconditioner combined with PINO-based warm-start initialization. 

The discretized PSFD system is written as
$\mathbf{A}\mathbf{E}=\mathbf{b}$.
Convergence is then assessed using the relative residual, which provides a dimensionless, amplitude-invariant measure:
\begin{align}
    r_{\mathrm{rel}}^{(k)}
    =
    \frac{
        \left\|
            \mathbf{b}
            -
            \mathbf{A}\mathbf{E}^{(k)}
        \right\|_2
    }{
        \left\|\mathbf{b}\right\|_2
    }.
    \label{eq:relative_residual}
\end{align}

\begin{figure}[H]
    \centering
    \begin{tikzpicture}
        \node[anchor=south west, inner sep=0] (image) at (0,0)
            {\includegraphics[width=1.0\textwidth]{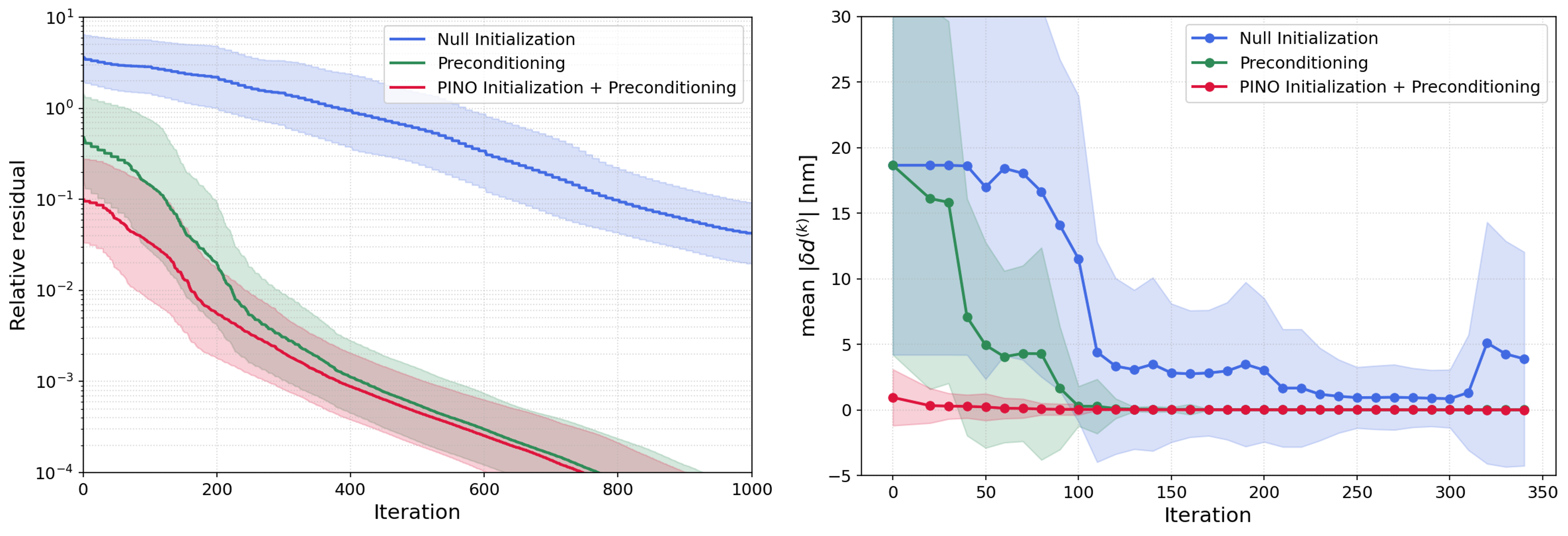}};
        \begin{scope}[shift={(image.south west)},
                      x={(image.south east)}, y={(image.north west)}]
            \node[anchor=north west, font=\bfseries, inner sep=1.5pt]
                at (0.01, 1.1) {(a)};
            \node[anchor=north west, font=\bfseries, inner sep=1.5pt]
                at (0.5, 1.1) {(b)};
        \end{scope}
    \end{tikzpicture}
    \caption{%
        (a) Relative residual as a function of iterations for the three solver configurations.
        (b) Mean and standard deviation of the wafer-equivalent near-field threshold-width deviation $|\delta d^{(k)}|$ over iterations for the three configurations.
    }
    \label{fig:figure5}
\end{figure}

\begin{figure}[H]
    \centering
    \begin{tikzpicture}
        \node[anchor=south west, inner sep=0] (image) at (0,0)
            {\includegraphics[width=1.0\textwidth]{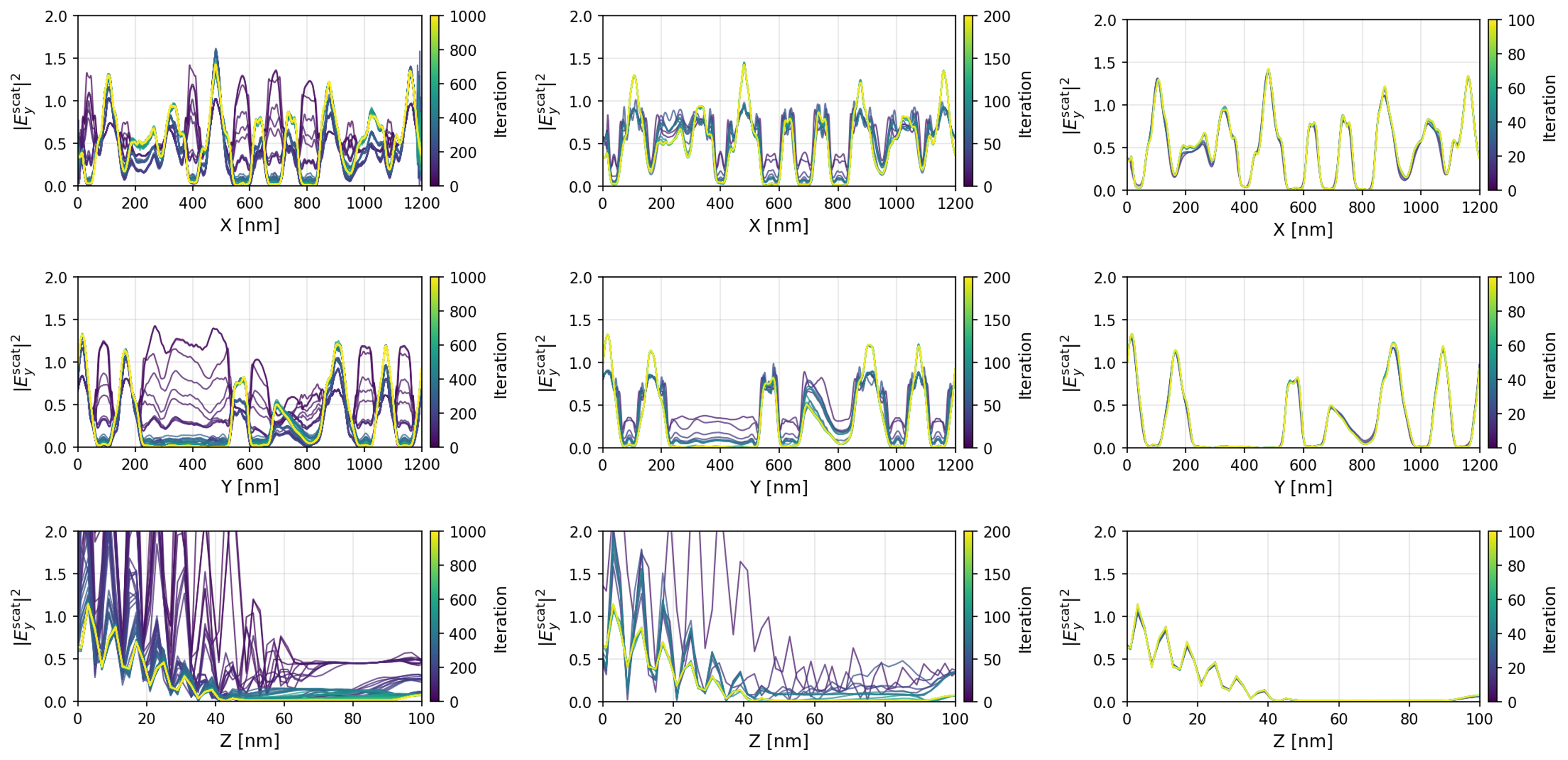}};
        \begin{scope}[shift={(image.south west)},
                      x={(image.south east)}, y={(image.north west)}]
            \node[anchor=north west, font=\bfseries, inner sep=1.5pt]
                at (0.01, 1.05) {(a)};
            \node[anchor=north west, font=\bfseries, inner sep=1.5pt]
                at (0.34, 1.05) {(b)};
            \node[anchor=north west, font=\bfseries, inner sep=1.5pt]
                at (0.67, 1.05) {(c)};
        \end{scope}
    \end{tikzpicture}
    \caption{%
        Convergence trends of the background-decomposed PSFD solver under three configurations, evaluated by the iteration-to-iteration change in the reflected scattered field $\mathbf{E}^{\mathrm{scat}}$ at the top surface of the absorber, along the x-, y- and z-axis indicated in \Cref{fig:figure3}.
        (a) Null initialization without preconditioning, shown up to 1,000 iterations. 
        (b) Scalar spectral-damping right preconditioning with null initialization, shown up to 200 iterations.
        (c) Scalar spectral-damping right preconditioning with PINO warm-start initialization, shown up to 100 iterations.
    }
    \label{fig:figure6}
\end{figure}
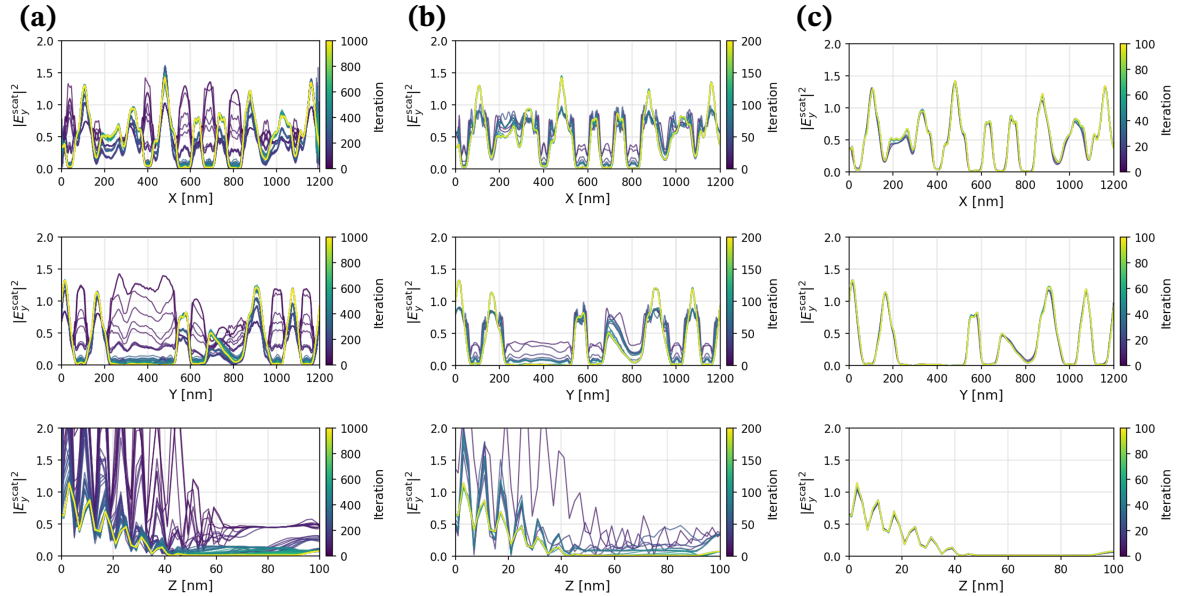

\Cref{fig:figure5}(a) plots $r_{\mathrm{rel}}^{(k)}$ as a function of iterations ($k$) for the three configurations.
In the null case, the relative residual decreases slowly with increasing iteration count, reaching approximately $5\times10^{-2}$ after 1,000 iterations. 
The scattered field progresses from a zero initial state toward an approximate solution over the course of 1,000 iterations. At this point, the dominant spectral components of the field have been largely recovered, as seen in \Cref{fig:figure6}(a). 
As shown later, this residual level represents a practically relevant operating point for rapid field computation, since the resulting scattered field shows reasonable accuracy for estimating variations in lithographically relevant feature dimensions even before fully converging.

Applying the scalar spectral-damping right preconditioner substantially steepens the observed residual decay, as shown in \Cref{fig:figure5}(a).
The spectral damping operator $\mathcal{M}$ is applied independently to each field component through a scalar Fourier multiplier:
\begin{align}
    \mathcal{F}_{3}\!\left[\mathcal{M}\mathbf{x}\right](\mathbf{k})
    =
    m(\mathbf{k})\,\mathcal{F}_{3}\!\left[\mathbf{x}\right](\mathbf{k}),
    \label{eq:spectral_damping_action}
\end{align}
where the three-dimensional Fourier transform is applied componentwise and
\begin{align}
    m(\mathbf{k})
    =
    m_{\min}
    +
    (1-m_{\min})
    \begin{cases}
        1,
        &
        |\mathbf{k}| \leq k_{\mathrm{ref}},
        \\[3pt]
        \left(
            \dfrac{k_{\mathrm{ref}}^2}
                  {|\mathbf{k}|^2+\epsilon}
        \right)^{\gamma},
        &
        |\mathbf{k}| > k_{\mathrm{ref}}.
    \end{cases}
    \qquad
    |\mathbf{k}|^2=k_x^2+k_y^2+k_z^2.
    \label{eq:spectral_damping_mask}
\end{align}
Here, $m_{\min}$ is the multiplier floor, $k_{\mathrm{ref}}$ is the cutoff wavenumber, $\gamma$ controls the high-frequency decay, and $\epsilon$ is a small numerical regularizer.
With $m_{\min}=0$, $\gamma=1$, $\epsilon\rightarrow0$, and $k_{\mathrm{ref}}=k_{\mathrm{bg}}$, Eq.~\eqref{eq:spectral_damping_mask} reduces to the isotropic spectral damping baseline $D(k)=\min(1,k_{\mathrm{bg}}^2/k^2)$ used in Ref.~\cite{kim2026psfd}.
The present form extends that baseline by introducing a nonzero floor and a tunable decay exponent.
With this approach, the Krylov solver solves for an auxiliary variable $\mathbf{x}$, from which the physical field $\mathbf{E}$ is recovered as 
\begin{align}
    \mathbf{A}\mathbf{E}
    =
    \mathbf{A}\mathcal{M}\mathbf{x}
    =
    \mathbf{b}
    \label{eq:right_preconditioned}
\end{align}

This makes the Krylov iteration more efficient, as the preconditioner $\mathcal{M}$ effectively conditions high-frequency components that are typically associated with slow convergence.
The PINO warm start extends this formulation by introducing a nonzero base field through an affine right-correction formulation.
Letting $\mathbf{E}_{\mathrm{base}}$ denote the initial field, the solution is written as
\begin{align}
    \mathbf{E}
    =
    \mathbf{E}_{\mathrm{base}}
    +
    \mathcal{M}\mathbf{y},
    \label{eq:right_correction_solution}
\end{align}
and the Krylov solver is applied to
\begin{align}
    \mathbf{A}\mathcal{M}\mathbf{y}
    =
    \mathbf{b}
    -
    \mathbf{A}\mathbf{E}_{\mathrm{base}}.
    \label{eq:right_preconditioned_system}
\end{align}


For null initialization, $\mathbf{E}_{\mathrm{base}}=\mathbf{0}$ and $\mathbf{y}=\mathbf{x}$, so the affine correction system reduces to the standard right-preconditioned formulation in Eq.~\eqref{eq:right_preconditioned}.
The warm-start field $\mathbf{E}_{\mathrm{base}}$ is therefore preserved, while the damping operator is applied only to the iterative correction $\mathbf{y}$. The scalar multiplier leaves low-spatial-frequency corrections unchanged and attenuates higher-frequency corrections down to the floor $m_{\min}$. This modification empirically improves the convergence behavior for the problems considered here.
At a relative residual of $r_{\mathrm{rel}}^{(k)} \approx 5\times10^{-2}$, where the MAE falls below $1\times10^{-3}$, the preconditioned-only configuration reaches this accuracy level in approximately 150 iterations, whereas the null configuration requires approximately 1,000 iterations.

When the PINO warm-start is additionally applied, the iteration begins from an approximately one-order-of-magnitude lower initial residual, shifting the entire decay curve downward. 
The same relative-residual level of $5\times10^{-2}$ is reached in approximately 75 iterations, corresponding to a factor-of-2 reduction relative to the preconditioned-only case and a factor-of-13 reduction relative to the null case.
The representative end-to-end wall-clock times are approximately 30 minutes, 4 minutes, and 2 minutes for the null, preconditioned-only, and PINO-warm-start configurations, respectively. 
For the PINO case, the reported time includes model loading, neural-operator inference, grid interpolation, and field conversion.
As shown in \Cref{fig:figure6}(c), the scattered field starts close to the final solution and changes only marginally over the course of the iteration, with the remaining updates confined to fine-scale refinements. This early transient phase is of primary practical interest, as the dominant field components are established here and practical convergence criteria may already be satisfied.

The preconditioned null-initialized curve exhibits a slightly steeper decay during part of the early transient than the PINO-initialized curve. 
This difference reflects the different spectral content and magnitudes of their initial residuals. The PINO warm start has already removed a substantial portion of the dominant initial field error, whereas the null-initialized solve begins with a larger residual distributed over a broader range of spectral components. 
As the iteration proceeds toward tighter tolerances, the two preconditioned
configurations exhibit increasingly similar empirical decay rates.
Accordingly, the benefit of PINO initialization becomes progressively less pronounced.

\paragraph*{Practical assessment of scattered-field convergence.}
Following the threshold-contour procedure used for CD extraction from aerial or resist images, we apply an analogous contour-based metric directly to the scattered-field intensity to assess a practically useful accuracy level without performing full projection-optics or resist simulation. 
For this, feature widths are extracted and evaluated from crossings of a fixed intensity threshold along horizontal and vertical scan lines.
For each segment bounded by two threshold crossings, we track the deviation $\delta d^{(k)}=d_{\mathrm{ref}}-d^{(k)}$, where $d^{(k)}$ is the threshold-contour width at iteration $k$.
The reference value $d_{\mathrm{ref}}$ is obtained from the fully converged PSFD solution at $r_{\mathrm{rel}}^{(k)} \approx 10^{-7}$.
This metric additionally quantifies the stability of the scattered near-field geometry as the iterative solution converges.

For comparison with wafer-scale lithographic dimensions, the measured contour displacements are divided by four, corresponding to the ideal isotropic $4\times$ demagnification of a 0.33-NA EUV optical system.
The metric does not include projection-optics filtering, partial coherence, aerial-image formation, or resist-process effects, and should therefore not be directly interpreted as a predicted printed-feature CD error.
The intensity field is mapped through the logistic sigmoid function
\begin{align}
    S(I)
    =
    \frac{1}{
        1 + \exp\!\left(
            -\frac{I-I_{\mathrm{th}}}{w}
        \right)
    },
    \label{eq:logistic_sigmoid}
\end{align}
where $I = |E_y^{\mathrm{scat}}|^2$, and $I_{\mathrm{th}} = 0.15$ and $w = 0.05$ are the threshold and transition-width parameters adopted in this study.


\begin{figure}[H]
    \centering
    \includegraphics[width=1\textwidth]{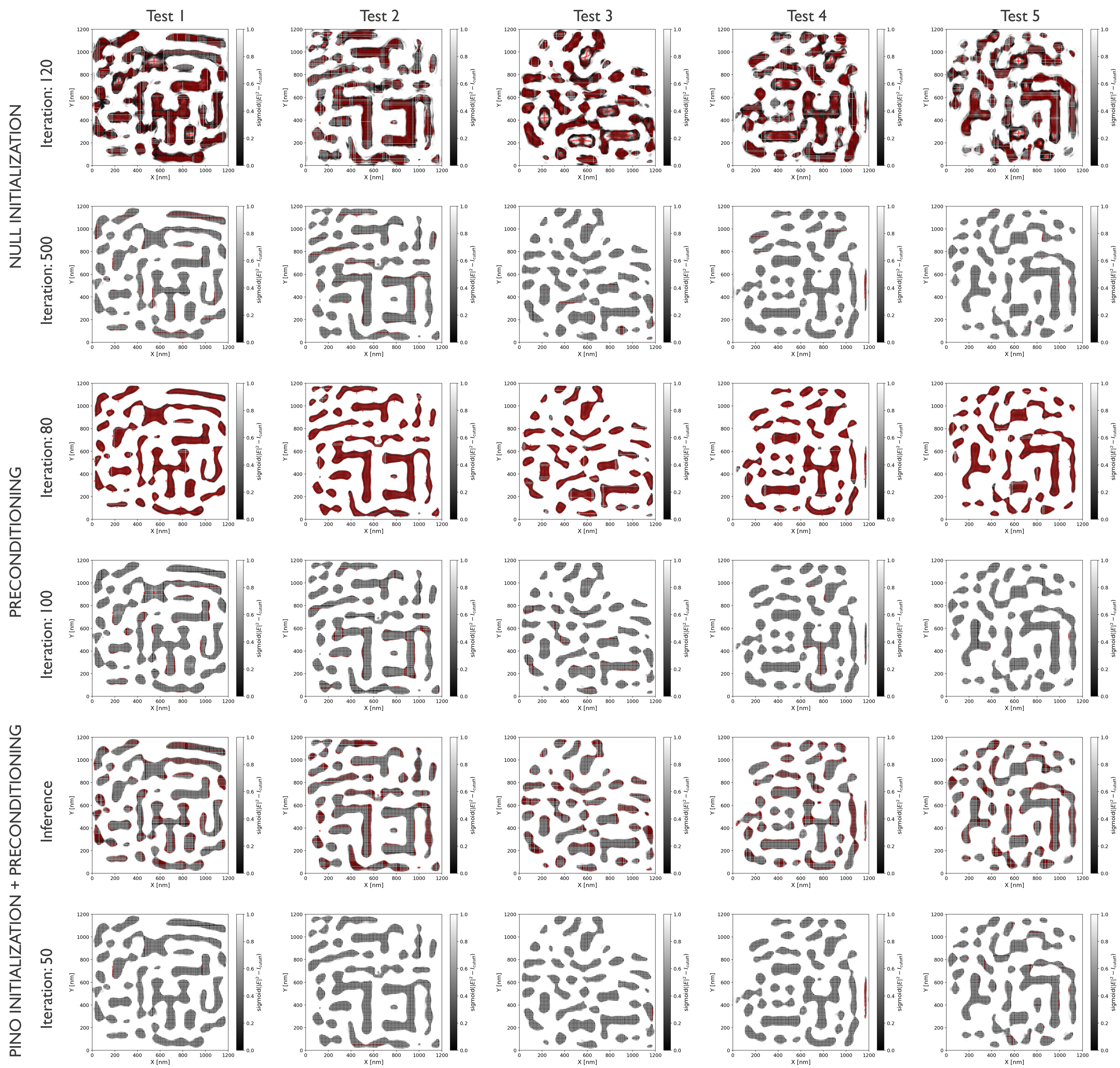}
    \caption{%
        Spatial maps of the wafer-equivalent near-field width error
        $|\delta d|$ overlaid on the thresholded scattered-intensity field
        $S(|E_y^{\mathrm{scat}}|^2)$ for the three solver configurations at
        representative early and late iterations.
        Red segments indicate scan-line widths with $|\delta d|>1\,\mathrm{nm}$,
        red crosses indicate locations where no resolvable threshold-bounded
        segment is formed, and white segments indicate
        $|\delta d|\leq1\,\mathrm{nm}$.
    }
    \label{fig:figure7}
\end{figure}

The threshold-contour width $d^{(k)}$ is extracted at the $S=0.5$ contour, which corresponds to $I=I_{\mathrm{th}}$.
At each iteration, $d^{(k)}$ is measured along 100 horizontal and 100 vertical scan lines per pattern and averaged over 100 test patterns not used during training. 
\Cref{fig:figure5}(b) shows the mean and standard deviation of $|\delta d^{(k)}|$ over iterations for the three configurations. In the null case, $|\delta d^{(k)}|$ is initially large and decreases only slowly. \Cref{fig:figure7} visualizes these trends by overlaying scan-line segments on the sigmoid-binarized intensity field at representative iteration snapshots. 
Scan-line positions where $|\delta d|>1\,\mathrm{nm}$ are shown in red, whereas those where $|\delta d|\leq1\,\mathrm{nm}$ are shown in white.
For the null configuration, above-threshold segments dominate the map at iteration 120, reflecting widespread above-threshold deviations, while by iteration 500 the fraction of above-threshold segments decreases significantly, confirming the slow but gradual approach to a sub-nanometer near-field threshold-width deviation in wafer-equivalent units.

The spectral damping preconditioner accelerates this decay markedly, 
reducing both the mean and standard deviation of the near-field
threshold-width proxy below $1\,\mathrm{nm}$ in wafer-equivalent
units within about 150 iterations.
The fraction of above-threshold segments in \Cref{fig:figure7} is dramatically reduced from iteration 80 to 100, confirming the observation from the relative residual that the preconditioner accelerates convergence in the early transient correction phase.

With additional PINO warm-start initialization, $|\delta d^{(k)}|$ is already relatively small from the inference stage and the fraction of above-threshold segments is correspondingly low. 
This method reaches the same sub-nanometer near-field proxy level within about 75 iterations. 
These results demonstrate that the PINO surrogate can serve as a warm-start initializer for the background-decomposed PSFD solver, reducing the number of iterations required to satisfy the prescribed practical convergence criterion. 
When strict convergence to $r_{\mathrm{rel}}^{(k)}$ of $10^{-7}$ is required, however, the required iterations increase dramatically, with the null case requiring about 11,000 iterations, and either preconditioned configuration requiring about 2,000 iterations.

\section{Conclusion}
We have presented a background-decomposed PSFD framework combined
with a PINO surrogate for scalable EUV mask simulation. In this framework, the computational domain is substantially reduced to the mask absorber region, while the multilayer response is incorporated analytically through a TMM-based reflection operator during both training and rigorous simulation.

At its native training resolution, the first PINO model achieves an MAE of approximately $7 \times 10^{-3}$ on held-out mask geometries.
For the finer-grid evaluation, a separately trained PINO model is applied without retraining to a four-times finer lateral discretization of the same physical domain, yielding an MAE of approximately $3 \times 10^{-2}$.
In this case, the error is largely concentrated near material interfaces, where the finer inference grid resolves spatial-frequency content beyond that represented on the coarser training grid.

The scalar spectral-damping right preconditioner reduces the iteration count by a factor of approximately 6.7 relative to the null case under the prescribed convergence criterion. 
When combined with the preconditioner, PINO warm-start initialization accelerates the early iteration phase, reducing the number of iterations required to reach a practically useful accuracy level by a factor of approximately 13 relative to the null baseline.

The corresponding representative end-to-end wall-clock times are about 30 minutes for null initialization and about 1--2 minutes for PINO warm-start initialization, with the latter including model loading, inference, interpolation, and field conversion. 
When tight convergence is required, the two preconditioned configurations reach the target with similar iteration counts, and the benefit associated with PINO initialization becomes less pronounced. 
The proposed framework provides an effective route to rapid large-scale field computation by combining fast PINO-based initialization and preconditioning with the accuracy of the rigorous PSFD solver.

\section{Acknowledgments}
The authors express gratitude to Seungjin Lee, Pervaiz Kareem, Hazem Mesilhy and Xuelong Shi for valuable discussions during the initial stages of this work. This work has been enabled in part by the NanoIC pilot line. The acquisition and operation are jointly funded by the Chips Joint Undertaking, through the European Union's Digital Europe (101183266) and Horizon Europe programs (101183277), as well as by the participating states Belgium (Flanders), France, Germany, Finland, Ireland and Romania. For more information, visit nanoic-project.eu.

\bibliographystyle{jhep}
\bibliography{psfd}
\end{document}